\documentstyle[preprint,aps,eqsecnum,tighten]{revtex}
\preprint{\vbox{\baselineskip=12pt
\rightline{CGPG-95/8-5}
\rightline{gr-qc/9508037}}}

\begin{document}
\draft
\title{Quantization of Nonstandard Hamiltonian Systems}
\author {Alejandro Corichi\thanks{Electronic address:
 corichi@phys.psu.edu} and
 Michael P. Ryan Jr.\thanks{Electronic address: ryan@phys.psu.edu}
\thanks{Permanent Address:
Instituto de Ciencias Nucleares, UNAM, A. Postal 70-543,
 M\'exico 04510 D.F.,
MEXICO.}}
\address{Center for Gravitational Physics and Geometry \\
Department of Physics, The Pennsylvania State University \\
University Park, PA 16802}
\maketitle
\begin{abstract}

The quantization of classical theories that admit more than one
Hamiltonian description is considered.
This is done from a geometrical
viewpoint, both at the quantization level
 (geometric quantization)
and at the level of the dynamics of the quantum theory.
A spin-1/2 system is taken as an example in which all
 the steps can be completed.
It is shown that the geometry of the quantum theory imposes
restrictions on the physically
 allowed nonstandard quantum theories.

\end{abstract}
\pacs{PACS number(s): 03.65.Bz, 03.65.Ca, 02.40.-k}

%\narrowtext

\section{Introduction}

The problem of quantization of a classical theory is at least
 seventy years old,
but the term `quantization' always has had a somewhat
loose meaning. There is no such thing a {\it the\/} quantization
 recipe that takes
a classical theory and produces for us the `correct' quantum theory.

There are three main approaches to canonical quantization:
 algebraic \cite{algq},
 geometric \cite{wood}, and group theoretic quantization \cite{groupq}.
They differ,
roughly speaking, in the basic structures on phase space they regard as
 fundamental in order to construct a quantum theory.
 In each of these approaches one is led to
make several choices along the way that might yield inequivalent quantum
 theories. Well known examples of
 these ambiguities are the factor ordering
 problem and different representations of the CCR in QFT, for example.

The quantization schemes mentioned above have, however,
a common feature.
They assume that the classical system to be
quantized is unique, that is, that there is a preferred classical
description for the system.  From the classical viewpoint,
 on the other hand,
there might be more than one perfectly valid
 way of representing a given system. These alternative
descriptions are called {\it nonstandard\/} Hamiltonian systems.
 The aim of this paper is
to explore the possibility of quantization
 starting from  different classical theories.

The program of quantization of nonstandard Hamiltonian
 dynamics has its roots in work of Feynman
reported by Dyson \cite{dyson} and its extension
 by Hojman and Shepley \cite{Hojshep}.
Feynman's original work showed that
Poisson-bracket relations place
strong constraints on the types of forces allowed
in physical systems.
Hojman and Shepley generalized Feynman's work and were able to
show that a consistent quantization with
 a set of commuting coordinates
led to a second order Lagrangian in those coordinates.
Hojman then constructed a consistent Poisson-bracket Hamiltonian
 theory for first-order equations of motion of the
form $\dot x^i = f^i (x^j)$ \cite{hoj1}.  We will discuss
 this formalism
 in more  detail below. The question was open, however,
about the possibility of
quantizing those systems that admit no Lagrangian.

This program could be seen as yet another
ambiguity in the quantization
process or, if viewed from a different
 perspective, as a new avenue for
finding  possibly valid quantum theories.
This would be the case, for
 instance, if
the given system has more than one classical description without any
a priori criteria for choosing the `correct' one.

We will proceed as follows. In the introduction we will recall the
 basic steps
of geometric quantization, pointing out the choices one makes in the
process and
discussing the possible implications in the final quantum theory.
Section~\ref{sec:sec2} reviews the possibility of
 different classical descriptions or
`non-standard Hamiltonian systems'.
 We consider as an example the classical
spinning particle.
 Section~\ref{sec:sec3} recalls the geometry of quantum mechanics as
proposed by
Ashtekar and Schilling, focusing in the spin 1/2 particle.
The basic program is discussed in Sec.~\ref{sec:sec4} for the spinning
particle.
 The obstructions to quantizing
the nonstandard description are isolated.
 Section~\ref{sec:sec5} discusses the results
 and suggests some directions for further research.
Throughout the paper, the `abstract index notation' is employed.
 For a discussion of the notation see \cite{wald}.

\paragraph{Geometric Quantization.}
By quantization we will mean the process
of finding a quantum theory from some known classical theory.
 The starting point for all
canonical quantization schemes is a
classical system described in terms of symplectic
geometry. Let us recall the basic notions
 in order to set the notation \cite{{arnold},{marsden}}.

The {\it phase space\/} of the system
 consists of a manifold  $\Gamma$ of
dimension $dim (\Gamma )=2n$ (real).
 Physical states are represented by
the points on the manifold.
 {\it Observables\/} are smooth functions  on $\Gamma$.
There is a non-degenerate, closed
two-form $\Omega$ defined on it. That is,
the form $\Omega _{ab}$
satisfies $\nabla_{[c}\Omega _{ab]}=0$, and if
 $\Omega _{ab}V^b=0$ then
$V^b=0$. Therefore, there exists an inverse
$\Omega ^{ab}$ which defines
an isomorphism between the cotangent and the
 tangent space at each point of
$\Gamma$. The existance of the  {\it symplectic two-form\/}
$\Omega$ endows $(\Gamma, \Omega)$ with a
 {\it symplectic structure}.

A vector field $V^a$ generates infinitesimal
 canonical transformations if it
Lie drags the symplectic form, i.e.,
\begin{equation}
{\cal L}_V\Omega =0.
\end{equation}
This condition is equivalent to saying
that locally it is of the form:
$V^b=\Omega^{ba}\nabla_a f:= X^b_f$
and it is called the {\it Hamiltonian vector field of $f$ (w.r.t.
 $\Omega$)}.
Note that the symplectic structure gives us a mapping between
functions on
$\Gamma$ and Hamiltonian vector fields.
Thus, functions on phase space (i.e. observables)
are generators of infinitesimal canonical transformations.

The Lie algebra of vector fields induces
a Lie algebra structure on the
space of functions.
\begin{equation}
\{ f,g\} :=\Omega_{ab} X^a_g X^b_f = \Omega^{ab}\nabla_af\nabla_bg,
\end{equation}
such that
$X^a_{\{ f,g\} }= -[X_f,X_g]^a$.

Since the symplectic form is closed,
 it can be obtained locally from a
{\it symplectic potential}, $\omega_a$,
\begin{equation}
\Omega_{ab}=2\nabla_{[a}\omega_{b]}.
\end{equation}

Time evolution is given by a vector field $f^a$ whose integral curves
are the dynamical trajectories of the system. On phase space there is a
{\it preferred\/} function, the {\it Hamiltonian\/}
 $H$ whose Hamiltonian vector
field corresponds directly to $f^a$, i.e.,
\begin{equation}
f^a=\Omega ^{ab}\nabla_b H \label{hameq}.
\end{equation}
Adopting the viewpoint that all observables generate canonical
 transformations
we see that the motion generated by the
 Hamiltonian corresponds to `time
evolution'. The `change' in time of the observables
 will be simply given
by the Poisson bracket of the observable with $H$
($\dot g=f^a\nabla_a g=\Omega^{ab}\nabla_a g \nabla_b H=\{ g, H\}$).

So far, not very much has been assumed about the phase space $\Gamma$.
 It can be any (even dimensional) manifold with complicated topology,
 compact, open, etc. The
symplectic structure $\Omega$ and the function $H$ are assumed to be
 given a priori.
Note that they might not be unique. From the classical
viewpoint the only
`observable' entities are the dynamical
 trajectories $f^a$ of the system
(the equations of motion). They could have
 come from more than one pair
 $(\Omega, H)$\footnote{There is another,
 even more drastic, possibility.
 There could
be another $f^{\prime a}$ that could have the same integral curves
as $f^a$. Such systems are called S-equivalent \cite{s-eq}.
We will not consider them here.}.

However, if the system has a configuration space $C$,
 then the phase space is
automatically `chosen' to be the cotangent bundle of
 the configuration space
$T^*C$. There is also a preferred
1-form on $C$ that can be lifted to $T^*C$
and taken to be the symplectic potential
which determines uniquely the symplectic
structure. Therefore, the fact that there exists a configuration
 space picks out for us the phase space
 and the symplectic two-form.

The program of quantization can be
divided in two parts: kinematical and dynamical.
The kinematical part deals with the problem of
defining  a good prescription for going `from Poisson brackets to
commutators' in a consistent way. That is,
 it should start with the classical
system and produce a Hilbert space of states.
The dynamical part deals with the
Hamiltonian, i.e. the generator of dynamical evolution.

We will concentrate on geometric quantization whose starting point
is  a symplectic manifold $(\Gamma, \Omega)$. There is no
{\it a priori\/} assumption about the structure
 of the phase space $\Gamma$.
It can be completely general.
In particular it can include the case in which $\Gamma$
 is compact, i.e., it is {\it not\/} a cotangent bundle.

There are two steps in geometric quantization.
 The first one involves defining
a Hilbert space on the full phase space.
 Wave functions are, roughly speaking,
functions on $\Gamma$. Any observable can be `quantized'.
 The second step
involves introducing an additional structure on $\Gamma$,
 a {\it polarization\/}
that will select those wave functions that
 depend only on `half of the
coordinates'. Physical observables are
 those that respect, in a way to be
defined below, the polarization.

We start with a Hamiltonian system as defined above. We define
what are called {\it prequantum wave functions}.
 They are cross sections
$\Psi$ of a complex line  bundle over $\Gamma$. The corresponding
$U(1)$ connection is the symplectic potential $\omega_a$
whose curvature is the
symplectic two form $\Omega_{ab}$. For each trivialization
 $\omega_a$ there corresponds
a function $\Psi_{\omega}$. If we change $\omega$ by
 a gauge transformation
$\omega_a \rightarrow \omega_a +\nabla _a g$ then
\begin{equation}
\Psi_{\omega^\prime}=e^{ig/\hbar}\Psi_{\omega}.
\end{equation}
There is a Hermitian inner product in this complex
vector space given by
the Liouville measure on $\Gamma$.
 The pre-Hilbert space would be the
completion with respect to this inner product.

Any observable $f$ ($f: \Gamma \rightarrow R$) has a corresponding
symmetric operator $O_f$ defined by:
\begin{equation}
O_f \Psi =\frac{\hbar}{i}X^a_f \nabla_a \Psi + f\,\Psi :=
\frac{\hbar}{i} X^a_f \left(\partial _a - \frac{i}{\hbar}
\omega _a \right) \Psi  + f\,\Psi.
\end{equation}
These operators are: i)linear; ii) gauge-covariant, iii) symmetric
(formally self-adjoint).

The assignment $f \rightarrow O_f$ is one
 to one and preserves the natural
Lie algebra structure,
\begin{equation}
[O_f,O_g]=-i\hbar O_{\{ f,g\} }\label{commrel},
\end{equation}
that is, one can assign a consistent operator to all observables.

It is known that
`actual' quantum wave functions depend only
 on `half' of the variables.
 We have to
`split' $\Gamma$ into two parts. This is done by choosing a
 {\it polarization\/}
$P$. It assigns at each point $\gamma$ a maximal
 subspace $P|_\gamma$ of
the complexified tangent space such that:

i) $V^a$ and $W^a \in P|_\gamma$ then $[V,W]^a \in P|_\gamma$ for all
$\gamma$

ii)  for all $V^a,W^a \in P$ then $\Omega_{ab}V^aW^b =0$
 for all $\gamma$.

If $P$ is real we have a `real polarization'.
The first condition implies that through each point
of $\Gamma$ there passes an $n$-dimensional submanifold,
 which is tangent to
the subspace $P|_\gamma$.
 The phase space is  then foliated by $n$-dimensional submanifolds.
The second condition implies that the Poisson bracket of any two
coordinates of this submanifold vanishes.

Given a polarization, a {\it quantum wave function\/}
is a cross section $\Psi$ satisfying
\begin{equation}
V^a \nabla_a \Psi =0.
\end {equation}
For all $V^a \in P$. This is called the
 {\it polarization condition}.

This condition tells us that the wave function depends only on $n$
coordinates $q^i$ `in involution'(For instance, if we have a
configuration space $C$ with
coordinates $q^i$, the standard polarization is the
 `vertical polarization'
spanned by $\{ \frac{\partial}{\partial p_i} \}$. We have then that
$\{ q^i, q^j\}=0$.)

Classical observables whose pre-quantum operators become well defined
operators are {\it good observables}. The condition is,
\begin{equation}
[O_f, V^a\nabla_a] \Psi=0 .
\end{equation}
For all $V^a \in P$.
This can be written classically as $[X_f,V]^a \in P$ for all $V^a$
(${\cal L}_V \, X_f \in P$).
We say then that $X^a_f$ {\it preserves the
polarization $P$}.
In particular, the operators corresponding
 to the coordinates $q^i$
preserve the vertical polarization and
 therefore are good observables.

A special kind of complex  polarization is called K\"ahler.
 An almost complex structure is a
tensor field ${J_a}^b$ such that ${J_a}^b{J_b}^c=-{\delta_a}^c$, and
it is a canonical transformation:
 ${J_a}^b{J_c}^d\Omega_{bd}=\Omega_{ac}$.
Then,
\begin{equation}
g_{ab}:= \Omega_{ac}{J^c}_b\label{1.10}
\end{equation}
is symmetric, non-degenerate, positive definite metric.
 The triplet $(\Omega,
J,g)$ equips $\Gamma$ with an almost K\"ahler structure.
 We can construct
on the phase space a Hermitian (complex) inner product
whose real part
is given by $g$ and the imaginary part by $\Omega$, i.e
$(,)=\frac{1}{2}g(,)- \frac{i}{2}\Omega (,)$.

The tensor field $J$ has eigenvectors in the
 complexified tangent space.
 Let us decompose any (complexified) $V^a$ into two parts,
\begin{equation}
V^a_{\pm}:=\frac{1}{2}(V^a \mp i{J^a}_bV^b)
\end{equation}
where $V^a_+$ is an eigenvector of $J$ with eigenvalue $i$.
Let's choose
the vector space spanned by those eigenvectors.
 It is a $n$-dimensional
(complex) vector space, and $\Omega_{ab}V^a_+V^b_+=0$.
If the distribution is integrable (the manifold can be given complex
 charts), the polarization is called K\"ahler.

In this case the polarization condition,
 on the section of the Hermitian
line bundle, involves considering {\it holomorphic\/} sections. When the
phase space $\Gamma$ is compact it is necessary to have holomorphic
sections. This is relevant, for instance, for the quantization of spin
systems.

Note that prequantization is a purely kinematical step.
 It produces a
(nonphysical) Hilbert
space on $\Gamma$ and every observable is pre-quantizable.
There is no external
input [other that the original $(\Omega, H)$ pair].

The choice of polarization, on the other hand,
 has both kinematical and
dynamical content. It is kinematical because
 it singles out the physically
relevant
quantum states from the pre-quantum Hilbert space and defines what the
physically admissible observables are, namely those that preserve the
polarization. This choice has also dynamical implications since the
Hamiltonian might {\it not\/} be compatible with $P$. It is the choice of
polarization that might lead to inequivalent quantum theories.

\section{Nonstandard  Classical Theory}
\label{sec:sec2}

As we mentioned in the introduction, we are interested in considering
 systems that might have a nonstandard classical description.
 By this we mean systems that
admit more than one Hamiltonian formulation or
 systems that obey certain
equations of motion that do {\it not\/} come
from a variational principle.

This section has two parts.
 In the first  we review the nonstandard Hamiltonian systems
 mentioned above,
considering a generalization of the
 symplectic formalism, namely that of
Poisson structures on a manifold.
 The second part takes a spinning classical
particle as a particular example of a system that admits
 nonstandard descriptions.

\subsection{Poisson Structures and Non-standard Dynamics}
\label{sec:sec2a}

In the introduction we gave an overview
of the standard Hamiltonian
 dynamics in terms of a symplectic structure $\Omega_{ab}$.
 It is possible to define dynamics by introducing
a more general structure known as a
 {\it Poisson (bracket) structure}
\cite{{arnold},{marsden}}. It consists of
a bivector $\Pi^{ab}=\Pi^{[ab]}$ on $\Gamma$
satisfying the Jacobi identity:
\begin{equation}
\Pi^{c[d}\nabla_c\Pi^{ab]}=0.
\end{equation}
It defines naturally a `generalized' Poisson bracket between
 functions on $\Gamma$.
\begin{equation}
\{ f,g \}_{\Pi} := \Pi^{ab}\nabla_b f\nabla_a g.
\end{equation}
It also defines a mapping from functions to vector fields
\begin{equation}
\stackrel{\pi}{X^a_f} := \Pi^{ab} \nabla_b f.
\end{equation}

Note that $\Pi^{ab}$ might be degenerate,
 in which case there will be
 {\it Casimir functions}. For instance,
if $\nabla_a C$ is a degenerate `direction' of $\Pi^{ab}$
$(\Pi^{ab}\nabla_b C= 0)$, then
 $\{ f,C \}_{\Pi}\equiv 0 ,\;\; \forall f$. That is,
$C$ `commutes' with all functions on $\Gamma$.

In the case of a nondegenerate symplectic structure, its inverse
 $\Omega^{ab}$
defines (locally) an `almost' one to one mapping between functions
 and Hamiltonian
vector field, that is, two functions will
 define the same vector field
if they differ by, at most, a constant function.
On the other hand, for a degenerate
Poisson structure, given a Casimir function $C$,
then two functions $f$ and $g$
will define the same vector field  ${X^a_f}=\Pi^{ab}
 \nabla_b f$ if
$f = g + h(C)$ where $h(C)$ is {\it any\/}
 (differentiable) function of $C$.

Given a phase space $\Gamma$, the dynamical evolution of a system
 is given by the integral curves of a vector field $V^a$.
The vector field gives at each point
of $\Gamma$ a set of equations of motion for the system.
 If we choose some local
coordinates $x^{\mu}, \;\;\mu=1,\ldots 2n$, then the rate of change
 of each coordinate
$x^{\mu}$ is given by the Lie derivative of $x^{\mu}$ along $V^a$,
\begin{equation}
\dot x^{\mu}:={\cal L}_V x^i = V^a \nabla_a (x^{\mu})= V^{\mu}(x)
\end{equation}
Recall that in the $x^{\mu}$ coordinate system, $V^a= V^{\mu}(x)
 \left(\frac{\partial}{\partial x^{\mu}}\right)^a$.

A natural question is whether the given system of
 first order differential
equations
can be put in a Hamiltonian form. That is,
does there exist a Poisson structure
$\Pi^{ab}$ and a function $h$ such that $V^a=\Pi^{ab}\nabla_b h$?
If the set of equation came from a (second order) variational
 principle, then the Poisson
structure is the inverse of the (naturally defined)
 symplectic structure
${\Omega}^{(0)}_{ab}$ on $\Gamma=T^*C$ and the Hamiltonian $h$
is the Legendre transform of the
 Lagrangian (for non-singular systems).

There might be, however, {\it another\/} Poisson
 structure that makes the
 equations Hamiltonian, with another Hamiltonian.
 Those systems are  known as {\it bi}-Hamiltonian \cite{biham}.

In the case when the set of equations does
 not come from a variational principle,
there is in principle no natural way of putting
then in Hamiltonian form.
 A program
for doing this has been proposed in the past years by S.
 Hojman \cite{hoj1}.
The
underlying idea is that one should use the
 symmetries of the equations of
motion in order to construct a Poisson structure.
 Let us summarize the Hojman
construction for systems with $N=2n$
 constants of motion $C_i$, $(N - 1)$
 of which
do not depend explicitly on time. That is, one knows them as explicit
functions of the coordinates (a fairly strong requirement, equivalent
to knowing the full classical solution).  The preceding requirement is
sufficient to be able to reduce the equations to Hamiltonian form.  It
is, of course, not necessary for constructing the Hamiltonian theory.

This $\Pi^{ab}$ may be constructed by summing elements of the form
\begin{equation}
 \Pi^{ab} = \mu (x) \varepsilon^{ab \lambda_1 \cdots \lambda_{N - 2}}
\nabla_{\lambda_1}C_{1} \cdots \nabla_{\lambda_{N - 2}} C_{N - 2} ,
\label{2.5}
\end{equation}
where  $\varepsilon
^{ab \lambda_1 \cdots \lambda_{N - 2}}$ is the $N$-index Levi-Civita
symbol, and $\mu (x)$ is a function of the coordinates to be
explained below.  This $\Pi^{ab}$ satisfies the Jacobi identity.  The
$C_1, \cdots ,C_{N - 2}$ are time-independent
constants of motion.  The
Hamiltonian is defined by $H = C_{N - 1}$,
 along with $C_N = t + d_N$,
where $d_N$ is time-independent.
 This can always be achieved by a change
of coordinates. Hojman has another
 construction that uses a symmetry of the
equations of motion, without needing
to know some constants of the motion
in explicit form. For more details see \cite{hoj1}.

Suppose that for a given set of equations that come from a
Lagrangian, we have been able to construct a
 non-degenerate $\Pi$ by means
 of the Hojman procedure.
 Let us denote by $\Omega^{\prime}_{ab}$ the corresponding two-form
$(\Omega^{\prime}_{ab}\Pi^{bc}=\delta^c_a$). If the
Poisson structure $\Pi$ is compatible with $\Omega^{ab}$
\footnote{Two Poisson structures are said to be
{\it compatible\/} if their sum
is also a Poisson structure \cite{biham}.},
then there will be a tensor field $K^a_b$ such that
\begin{equation}
\Omega^{\prime}_{ab}=K^c_a \Omega_{cb}.\label{hoj}
\end{equation}
Note that since $\Omega$ is invertible, we have then
 $K^d_a=\Omega^{\prime}_{ab}\Omega^{bd}$. We will call this mapping a
 {\it Hojman transformation}.

\subsection{Classical Description of a Spin-1/2 particle}
\label{sec:sec2b}

As we mentioned in the Introduction, the example
 we would like to use to
illustrate the difficulties of changing Poisson structures in quantum
mechanics is the simplest quantum system, that of a spin-1/2 particle.
In order to investigate the relationship between the classical and
quantum theories we would like to study
the classical problem equivalent to
that of a quantum spin-1/2 particle.
The main difficulty with this idea is
that, strictly speaking, there is no classical limit to this problem.
There are a number of `classical' limits that have been proposed
\cite{corb}, but
we will use a limit in terms of Grassman variables.  We would like to
find a limit of the quantum theory based on the three spin operators
$\hat S_i = \hbar \sigma_i$, the $\sigma_i$ the Pauli matrices with
Hamiltonian $\hat H = A\hat S_3$, $A$ = const.
Notice that $\hat S_i^2 = \hbar^2$, and $[\hat S_i, \hat S_j
] =\hbar \varepsilon_{ijk} \hat S_k$, and
 $\{ \hat S_i, \hat S_j\}_+ = 0$,
$i \neq j$.  As $\hbar \rightarrow 0$, we get
 $\hat S_i^2 = 0$, $[\hat S_i,
\hat S_j] = 0$ and $\{\hat S_i, \hat S_j\}_+ = 0$,
 and there is no set of
classical numbers that can obey these relations.  If we write the
classical variables as $S_i = \varepsilon s_i (t)$,
 where the $s_i$ are
commuting functions of $t$ and $\varepsilon$ is a constant Grassman
number, then
$S_i^2 = 0$ ($\varepsilon^2 = 0$), $[S_i, S_j] = 0 = \{S_i, S_j\}_+$.

Assume we  have a Hamiltonian $H$,
 in principle a function of some coordinates $q_i$,
$i=1,2,3$,
and $S_i = \beta_{ik} p_k$, where the $p_i$ are the
 momenta conjugate to
the $q_i$, and $\beta_{ij} = \beta_{ij} (q)$
 (the angular velocities are
$\omega_i = \alpha_{ij} (q) \dot q_j$, where
$\alpha_{ij} \beta_{jk} =
\delta_{ik}$), then
\begin{equation}
\dot S_j + \gamma_{jk\ell} {{\partial H}\over
 {\partial S_k}} S_{\ell} = 0
\end{equation}
if $H$ does not depend explicitly on the $q_i$, {\it i.e.\/}, $H = H
(S_i)$.  For a rigid body $\gamma_{jk\ell} = \alpha_{\ell m} \left (
{{\partial \beta_{mk}}\over {\partial q_n}}\beta_{nj} - {{\partial
\beta_{mj}}\over {\partial q_n}} \beta_{nk}\right )$ $=
-\varepsilon_{jk\ell}$.  If we take $H = AS_3$ then
\begin{equation}
\dot S_i = \varepsilon_{i3k} AS_k,
\end{equation}
or,
\begin{equation}
\varepsilon \dot s_i = \varepsilon_{i3k} A \varepsilon s_k,
\end{equation}
and
\begin{equation}
\dot s_i = \varepsilon_{i3k} A s_k.
\end{equation}
These imply that $s_3 =$ const. $= K_1$ and
\begin{eqnarray}
\dot s_1 &=& -As_2, \label{2.10}\\
\dot s_2 &=& A s_1, \label{2.11}
\end{eqnarray}
so $s_1^2 + s_2^2 = $ const.
  These mean that $s^2_1 + s^2_2 + s_3^2 =
 {\cal S}^2=$ const.
 which implies that the classical state space is a two-sphere.
The system orbits lie on the  two-sphere of radius ${\cal S}$
and since $s_3$ is a constant they  are parallels of
`latitude'.
If we look at the equations for $s_i$, $\dot s_3 = 0$ and
(\ref{2.10},\ref{2.11}), they can be written as
\begin{equation}
\dot s_i = \Pi_{ij} {{\partial H}\over {\partial s_j}},
\end{equation}
with $H = As_3$ and $\Pi_{ij} = \varepsilon_{ijk} s_k$.

This is precisely an example of a very well studied
 system with a Poisson structure.
Systems that have rotational degrees of freedom
(a rigid body for example), have
a common description coming from the fact that the rotation
 group SO(3) acts on
the system, as we now recall \cite{{arnold},{marsden}}.
 The  phase space is given by a
 3-dimensional vector space (that we can identify with $R^3$)
 with coordinates $s_i$ (it is the dual of
 the Lie algebra so(3)). The Poisson structure is given by
\begin{equation}
\Pi_{ij} = {C^k}_{ij} s_k
\end{equation}
where $ {C^k}_{ij}=\delta^{kn}\varepsilon_{nij}$ are the structure
 constants of so(3).
It is clearly degenerate (any antisymmetric tensor field in an
odd dimensional
space is).
 Note however, that $\Pi_{ij}$ induces a nondegenerate
symplectic structure on each sphere of radius ${\cal S}$. $R^3$ is
 then foliated
by {\it leaves\/} of symplectic manifolds. Furthermore, the `natural'
 Casimir
function is $K_0=\frac{1}{2}\delta^{ij}s_is_j$ which is
clearly constant on each sphere.
All Hamiltonian vector fields generated by $\Pi$ are tangent to
 the spheres and
therefore leave the Casimir unchanged.

Note that  $\Pi$ can be written as
\begin{equation}
\Pi_{ij}=\varepsilon_{nij}\frac{\partial K_0}{\partial s_n}
\end{equation}
which is precisely of the form (\ref{2.5}).

A remark is in order. With our formalism we could recover the Euler
 equations
for a rigid body if we chose the Hamiltonian to be the kinetic energy
$T=I^{ij}s_is_j$, where $I^{ij}$ is the inverse of the inertia tensor.
The Hamiltonian we have chosen for our system $H=As_3$ is
therefore not the `kinetic' energy of a rigid body,
 but resembles more that of a
`point-like'
object that might interact with an external potential (a constant
 magnetic field, for example).

The idea now, in order to find different descriptions for the system,
is to use the Hojman prescription for different
 constants of the motion.
 We have the functions $K_1 = s_3$ and $K_2 = s^2_1 + s^2_2$.
 Following  Hojman \cite{hoj2} we can now take $C = C(K_1, K_2)$,
 any arbitrary function of $(K_1, K_2)$, and a new `Hamiltonian'
$H = H(K_1, K_2)$, also any function of $K_1$ and $K_2$, and define
\begin{equation}
\tilde{\Pi}_{ij} = \mu (s_{\ell}) \varepsilon_{ijk} {{\partial C}\over
 {\partial s_k}},
\end{equation}
We would like to have then the equations of motion for $s_i$  as
\begin{equation}
\dot s_i = \tilde{\Pi}_{ij} {{\partial H}\over {\partial s_j}}.
\end{equation}
We can have the same equations as before if we choose $\mu$
properly and $C$ and $H$ satisfy one condition.
 If we look at the $s_3$
equation we have
\begin{eqnarray}
\dot s_3 & = &\mu \left [ {{\partial C}\over {\partial
K_1}} {{\partial K_1}\over {\partial s_2}} +
 {{\partial C}\over {\partial
K_2}} {{\partial K_2}\over {\partial s_2}}\right ] \left [ {{\partial
H}\over {\partial K_1}} {{\partial K_1}\over {\partial s_1}} +
{{\partial
H}\over {\partial K_2}} {{\partial K_2}\over {\partial s_1}}\right ] -
\nonumber \\
& - &\mu  \left [ {{\partial C}\over {\partial
K_1}} {{\partial K_1}\over {\partial s_1}} + {{\partial C}\over
 {\partial
K_2}} {{\partial K_2}\over {\partial s_1}}\right ] \left [ {{\partial
H}\over {\partial K_1}} {{\partial K_1}\over {\partial s_2}} +
 {{\partial
H}\over {\partial K_2}} {{\partial K_2}\over {\partial s_2}}\right ],
\end{eqnarray}
and since $K_1$ does not depend on $s_1$ or $s_2$,
\begin{equation}
\dot s_3 = -2\mu s_1s_2 \left [ {{\partial C}\over {\partial K_2}}
{{\partial H}\over {\partial K_2}} - {{\partial C}\over
 {\partial K_2}}
{{\partial H}\over {\partial K_2}}\right ] = 0.
\end{equation}
For $s_1$
\begin{eqnarray}
\dot s_1 & = &\mu \left [ {{\partial C}\over {\partial
K_1}} {{\partial K_1}\over {\partial s_3}} + {{\partial C}
\over {\partial
K_2}} {{\partial K_2}\over {\partial s_3}}\right ] \left [ {{\partial
H}\over {\partial K_1}} {{\partial K_1}\over {\partial s_2}}
 + {{\partial
H}\over {\partial K_2}} {{\partial K_2}\over {\partial s_2}}\right ] -
\nonumber \\
& - & \mu  \left [ {{\partial C}\over {\partial
K_1}} {{\partial K_1}\over {\partial s_2}} + {{\partial C}\over
 {\partial
K_2}} {{\partial K_2}\over {\partial s_2}}\right ] \left [ {{\partial
H}\over {\partial K_1}} {{\partial K_1}\over {\partial s_3}} +
 {{\partial
H}\over {\partial K_2}} {{\partial K_2}\over {\partial s_3}}\right ],
 \nonumber \\
& = & 2\mu s_2 \left [{{\partial C}\over {\partial K_1}} {{\partial
H}\over {\partial K_2}} - {{\partial C}\over {\partial K_2}}
 {{\partial
H}\over {\partial K_1}} \right ].
\end{eqnarray}
We can achieve $\dot s_1 = -As_2$ if $\Delta \equiv {{\partial C}\over
{\partial K_1}}{{\partial H}\over {\partial K_2}} - {{\partial C}\over
{\partial K_2}}{{\partial H}\over {\partial K_1}} \neq 0$ and we take
$\mu = -\frac{A}{2\Delta}$.  It is easy to show that
 his choice of $\mu$
also gives
$\dot s_2 = As_1$, so we recover the original equations of motion.

As an example of this procedure, take
the normal Hamiltonian $H =
As_3$ and $C = s^2_1 + s^2_2$.
If we look at the plane $s_1 = 0$, the
orbits intersect the circle $s^2_3 + s^2_2 = 1$.
 The lines of constant
$s_3=H/A$ and $C$ are perpendicular straight
 lines that form a coordinate
grid over the half plane given by the $s_2s_3$-plane with $s_2 > 0$.
The sphere $s_1^2 + s_2^2 + s_3^2 = {\cal S}^2$
intersects this half plane in a semi-circle, and
any point on this semi-circle represents the initial
point of a possible orbit,
and if we rotate the semi-circle around the $s_3$-axis
 then a point on it
traces out a parallel of `latitude'.  In the rectangular
grid of $C$ and $H/A$ we can always specify
 this point by particular
values of $C$ and $H/A$.

Now, the equation for $s_i$ is
\begin{equation}
{{ds_i}\over {dt}} = \mu (s_{\ell}) \varepsilon_{ijk}
 {{\partial C}\over
{\partial s_k}}{{\partial H}\over {\partial s_j}}.
\end{equation}
Note that this has the form
\begin{equation}
{{d {\bf s}}\over {dt}} = \mu ({\bf s}) ({\bf \nabla} H) \times ({\bf
\nabla} C),
\end{equation}
where ${\bf \nabla}C$ and ${\bf \nabla}H$ are the
 two-dimensional
gradients of $C$ and $H$ which
are the the normals to the
coordinate curves.
We have ${\bf \nabla}H \times {\bf \nabla} C =$ $|{\bf \nabla} H
\times {\bf \nabla}C|{\bf e}_1$, where ${\bf e}_1$ is
 the unit vector in
the $s_1$ direction.  Since in the $s_1 = 0$ plane
\begin{equation}
{{ds_1}\over {dt}} = -As_2,
\end{equation}
we see that (15) gives this if we take $\mu = -As_2
 /|{\bf \nabla}H \times
{\bf \nabla} C|$.  From Ref.\cite{hoj2} we
 see that this $\mu$ works for
all $s_1$, $s_2$.

As long as they form a complete
 coordinate grid in the $s_2s_3$-plane,
any functions $C$ and $H$ can be used in the formulation.
  Note that if
${\bf \nabla} H$ is parallel to
 ${\bf \nabla} C$ at any point (or the
norm of one of the vectors is zero), $\mu$
blows up.  This is the condition in Ref. \cite{hoj2}
 for the nonexistence of
$\mu$.  Notice also that $H$ is no longer the energy.

Let us now try to understand what we are doing from
a geometrical viewpoint.
The fact that we are using a preferred
 function (the Casimir) to define the
Poisson structure means that one-forms $w_a$ `transverse' to the
$C={\rm constant}$
surfaces are precisely the degenerate directions of $\Pi$.
 Hamiltonian vector fields
are always tangent to the surfaces and therefore
the motion they generate lies
within them. In the standard case of the rigid body, for example,
 the surfaces on
which the Casimir is constant are spheres precisely because they are
 the orbits of
the rotation group (coadjoint action on the dual
 of the Lie algebra) on $R^3$.
The symplectic structure induced on the
 spheres from the Poisson structure
on $R^3$ is precisely ($1/{\cal S}$ times) the area element
(Recall that any nondegenerate two-form
on a surface is proportional, by means of a conformal factor,
 to the area element).

Suppose that we now define a new Poisson
structure via a function whose
surfaces of constant value are not spheres but some `ellipsoids' (with
rotational symmetry around the $s_3$ axis). Now,
 the surfaces will not be
the orbits of the rotation group in 3 dimensions
 (see \cite{hoj2} for a
particular choice in which the resulting deformed
 algebra is SU(2)${}_q$). The change in
the induced symplectic structure, the `Hojman transformation',
 will be a simple conformal transformation.
We can conclude then that by a  rescaling of the
symplectic structure and
a corresponding change in the Hamiltonian,
 we have an infinite number of
classical descriptions for the system.

As we mentioned above, we would now like to
 apply the idea of changing the
symplectic structure to quantum mechanics.   In the next section
we will discuss this formulation and
 its extension to `K\"ahler quantum
mechanics' in the context of the
 spin-1/2 example outlined above.  We will
see that two obstructions exist to doing this in
the most simple-minded way.
These are both related to the fact that we need to define a probability
structure on the quantum-mechanical phase space.
 Probability structures are
often given in terms of linear operators on a Hilbert space.
  We will see
that both the definition of probabilities in
 `K\"ahler quantum  mechanics'
and the realization of dynamical quantities as linear operators place
strong constraints on the possible symplectic
 structures that are allowed.

\section{Quantum Mechanics}
\label{sec:sec3}

The question we want to address in this paper is the possible
 quantization of
systems that admit non-standard descriptions.
 If the system admits more that
one classical description, we are led to ask whether
 the quantum theories
are equivalent. If not, what are the criteria to choose the
 `correct' classical
description?

As we mentioned in the introduction, there are,
roughly speaking, two different
sets of issues about the quantum mechanics one has to address:
 kinematical and dynamical. The kinematical conditions, so to speak,
that the constructed quantum theory should satisfy, are mainly
related to the Heisenberg uncertainty relations.
 Commuting quantum observables can, in principle, be  simultaneously
measured. Such  quantum  observables correspond to
classical observables
that have vanishing Poisson brackets among them.
 Therefore, there is in
 principle
a way of distinguishing between, for instance, two different Poisson
structures. If the Poisson structure
in the classical theory is degenerate,
there will be Casimir functions and, therefore,
corresponding quantum Casimir operators.
 This will lead to `super-selected'
sectors that  should be detected experimentally.

There are another set of issues one has to consider
 when analyzing the
 dynamical
content of the theory. Quantum mechanics is a
 theory of measurement. If the
theory is to pass the test of `validity',
it should provide probabilities
for measuring eigenvalues of various operators as functions in time,
 that should be compatible with measurements.
 This is a condition to be
satisfied by  the dynamical evolution of the
 quantum system. This condition
is analogous to the corresponding classical
 condition that the dynamical
 evolution should be the integral curves of
 a preferred vector field.
This `dynamical condition' has a very clean
 geometrical  formulation when
quantum mechanics is cast in geometric language.

\subsection{Geometry of Quantum Mechanics}
\label{sec:sec3a}

 Quantum mechanics, with all its
postulates, can be put into geometric language.
In this subsection we will recall the geometry of
 quantum mechanics. For
details see \cite{{GQM1,GQM2}}.

The description we will give is for systems with a finite
dimensional Hilbert space but the generalization to the
infinite dimensional case is straightforward \cite{GQM2}. Denote by
 ${\cal P}$ the
space of rays in the Hilbert space ${\cal H}$.
 In this case ${\cal P}$
will be the complex projective space $CP^n$, since ${\cal H}$ can be
identified with $C^n$.

It is convenient to view ${\cal H}$ as a
{\it real\/} vector space equipped
with a complex structure (recall that a
complex structure $J$ is a linear
mapping $J:{\cal H} \rightarrow {\cal H}$
 such that $J^2=-1$). Let us
decompose the Hermitian inner product into
 real and imaginary parts,
\begin{equation}
\langle \Psi|\Phi\rangle =\frac{1}{2} G(\Psi ,\Phi) -\frac{i}{2}
\Omega(\Psi ,\Phi),
\end{equation}
where $G$ is a Riemannian inner product on ${\cal H}$ and $\Omega$ is
a symplectic form.

Let us restrict our attention to the sphere $S$ of normalized states.
The true space of states is given by the quotient of $S$ by the $U(1)$
action of states the differ by a `phase', i.e. the projective space
${\cal P}$. The complex structure $J$ is the generator of the $U(1)$
action ($J$ plays the role of the imaginary unit $i$ when the Hilbert
space is taken to be real). Since the phase rotations preserve the norm
of the states, both the real and imaginary
 parts of the inner product can
be projected down to ${\cal P}$.

Therefore, the structure on ${\cal P}$ which
is induced by the Hermitian
inner product is given by  a Riemannian metric $g$ and a symplectic
two-form ${\bf \Omega}$. The pair $(g,{\bf \Omega})$ defines a K\"ahler
structure on ${\cal P}$ (Recall that a K\"ahler structure is a triplet
$(M,g,{\bf \Omega})$ where $M$ is a
 complex manifold (with complex structure
$J$), $g$ is a Riemannian metric and
 ${\bf \Omega}$ is a symplectic two-form,
such that they are compatible).

The space ${\cal P}$ of quantum states has then
 the structure of a K\"ahler
manifold, so, in particular, it is a
 symplectic manifold and can be regarded
as a `phase space' by itself. It turns out that
 the quantum dynamics can
be described by a `classical dynamics', that is,
 with the same symplectic
description that is used for classical mechanics.
Let us see how it works. In quantum mechanics,
 Hermitian operators on
${\cal H}$  are
generators of unitary transformations (through
 exponentiation) whereas in
classical mechanics, generators of canonical
 transformations are real valued
functions $f\,: {\cal P} \rightarrow R$.
 We would like then to associate with
each operator $F$ on  ${\cal H}$ a function $f$
on ${\cal P}$. There is
a natural candidate for such function:
 $f:= \langle F\rangle|_S$ (denote it
 by $f=\langle F\rangle$).
The Hamiltonian vector field $X_f$ of
 such a function is a Killing field of
the Riemannian metric $g$. The converse
also holds, so there is a one to one
correspondence between self-adjoint
operators on ${\cal H}$ and real valued
functions (`quantum observables') on ${\cal P}$
whose Hamiltonian vector fields are
symmetries of the K\"ahler structure.

There is also a simple relation between a natural
 vector field on ${\cal H}$
generated by $F$ and the Hamiltonian vector field associated to
 $f$ on ${\cal P}$.
Consider on $S$ a `point' $\psi$ and an operator
 $F$ on ${\cal H}$.
 Define the vector
$X_F|_\psi:=\frac{d}{dt} \exp[-JFt]\psi|_{t=0}=-JF\psi$.
 This is the generator of
a one parameter family (labeled by $t$) of unitary
 transformation on ${\cal H}$. Therefore, it preserves
the Hermitian inner-product. The key result is that
 $X_F$ projects down to
${\cal P}$ and the projection is
 precisely the Hamiltonian vector field
$X_f$ of $f$ on the symplectic manifold
 $({\cal P}, {\bf \Omega})$.

Dynamical evolution is generated by the Hamiltonian
 vector field $X_h$ when we
choose as our observable the Hamiltonian $h=\langle H\rangle$.
 Thus, Schr\"odinger evolution
is described by Hamiltonian dynamics,
 exactly as in classical mechanics.

One can define the Poisson bracket between a pair of
 observables $(f, g)$ from
the inverse of the symplectic two form ${\bf \Omega}^{ab}$,
\begin{equation}
\{ f, g\} := {\bf \Omega}(X_g, X_f) = {\bf
 \Omega}^{ab}(\partial_af)(\partial_bg).
\end{equation}
The Poisson bracket is well defined for arbitrary
functions on ${\cal P}$,
but when restricted to observables, we have,
\begin{equation}
\langle -i[F,G]\rangle = \{ f,g \} .
\end{equation}
This is in fact a slight generalization of
 Ehrenfest theorem, since when we
consider the `time evolution' of the observable
$f$ we have  the Poisson bracket  $\{ f, h\}=\dot{f}$,
\begin{equation}
\dot{f}=\langle-i[F,H]\rangle.
\end{equation}

We have seen that the symplectic aspect
 of the quantum state space
 is completely analogous to classical mechanics.
Notice that, since only those functions whose Hamiltonian vector
 fields preserve the
metric are regarded as `quantum observables' on ${\cal P}$,
 they represent a very small
subset of the set of functions on ${\cal P}$.

There is another facet of the quantum state space ${\cal P}$
 that is absent in
classical mechanics: Riemannian geometry.
Roughly speaking, the information
contained in the metric $g$ has to do
with those features which are unique
to the quantum description, namely,
those related to measurement and
 `probabilities'.
We can define a Riemannian product $(f,g)$
 between two observables as
\begin{equation}
(f,g):= g(X_f,X_g)= g^{ab}(\partial_a f)(\partial_b g).
\end{equation}
This product has a very direct physical interpretation in terms
 of the dispersion
of the operator in the given state:
\begin{equation}
(f,f) = 2 (\Delta F)^2.
\end{equation}
Therefore, the length of $X_f$ is the
 uncertainty of the observable $F$.

The metric $g$ has also an important
 role in those issues related to
 measurements.
Note that eigenvectors of the Hermitian
 operator $F$ associated
to the quantum observable $f$ correspond
 to points $\phi_i$ in ${\cal P}$
at which $f$ has local extrema.
 These points correspond to zeros of the
Hamiltonian vector field $X_f$, and
 the eigenvalues $f_i$ are the values of
the observable  $f_i=f(\phi_i)$ at these points.

If the system is in the state $\Psi$,
what are the probabilities of
measuring the eigenvalues $f_i$? The answer
 is strikingly simple:
measure the geodesic distance given by $g$ from
 the point $\Psi$ to the
point $\phi_i$ (denote it by $d(\Psi,\phi_i)$).
 The probability of measuring
$f_i$ is then,
\begin{equation}
P_i(\Psi) = \cos^2\left[\frac{d(\Psi,\phi_i)}
{\sqrt{2}}\right].\label{3.7}
\end{equation}
Therefore, a state $\Psi$  is more likely to
`collapse' to a nearby state
than to a distant one when a measurement is
 performed. We will now turn our
attention to spin systems and in particular
 the quantum theory of a spin-1/2
particle.

\subsection{The Spin-1/2 System}
\label{sec:sec3b}

In this part we will find the quantum theory of
a spin-1/2 particle starting from
the classical description of Sec.~\ref{sec:sec2}.
 We will then discuss the quantum theory
in the geometric language just described.

\subsubsection{Geometric Quantization of Spin Systems}

In Sec.~\ref{sec:sec2}, we arrived at a kinematical
 description for systems with
`rotational degrees of freedom', that includes spin systems.
 We saw that the physically relevant space is $R^3$ that
 is foliated by spheres of radius ${\cal S}$.
That is, for each  value of  ${\cal S}$ we have
  a sphere which corresponds to the
reduced phase space of a particle with classical
 `intrinsic angular momentum'
equal to  ${\cal S}$. Since each sphere is a symplectic
manifold with a perfectly
defined symplectic structure on it, we can
employ the machinery of geometric
quantization that was outlined in the introduction.

We have then, $\Gamma= S^2$, $\Omega_{ab}=
 {\cal S}\, \sin\theta \nabla_{[a}
\phi\nabla_{b]}\theta$, where we have chosen
 spherical coordinates  $(\theta, \phi)$ for the sphere.
Note that the symplectic two form is $1/{\cal S}$ times the
area element of a sphere of radius  ${\cal S}$.

The first step in geometric quantization is to construct
 the pre-quantum line bundle.
There are, however, some integrality
 conditions that must be satisfied so that
the pre-quantum line bundle exists.
These conditions are the generalization of the
Bohr-Sommerfeld  quantum conditions:
\begin{equation}
\frac{1}{2\pi \hbar}\int_{S^2}\Omega = k ,
\end{equation}
where $k$ is an integer. Since $\int_{S^2}\Omega=4\pi \,{\cal S}$,
the condition reads ${\cal S}=\frac{\hbar}{2} k$.
 This is precisely the quantization of spin!
What this condition is telling us is that the
only symplectic manifolds that can be
quantized are those that correspond to classical systems whose
angular momentum is an integer multiple of $\frac{\hbar}{2}$.

The next step is to find a polarization in
 the phase space $\Gamma$.
 Note that the sphere $S^2$ is a compact
 manifold and therefore does
 not correspond to a cotangent bundle.
Luckily the sphere is a complex manifold and therefore admits
 a K\"ahler structure.
We can coordinatize it by $z$ (recall that
 the Riemman sphere is the complex
plane with the point at infinity).
 the symplectic two form reads then,
\begin{equation}
\Omega=i \,k \hbar\, \frac{dz\wedge d\bar{z}}{(1+z\bar{z})^2}.
\end{equation}

The Hilbert space of states will correspond
 then to holomorphic sections of a
complex line bundle over the sphere.
 A standard theorem in complex analysis
shows that the space of such sections is
{\it finite\/} dimensional. Furthermore,
holomorphic functions on the coordinate
 $z$ can be represented by,
\begin{equation}
\Psi(z)=\sum^k_{l=0} \left(\begin{array}{c}
 k \\ l \end{array}\right) \psi_l
z^l,
\end{equation}
where $\psi_l$ are constants.
In this way, one gets all the finite-dimensional,
 unitary, irreducible representations of SU(2).

Since we are interested in the spin 1/2
representation, we have to consider the
$k=1$ case, that is, the `smallest' quantizable sphere.
 The Hilbert space
in this case is given by elements of the form,
\begin{equation}
\Psi=\psi_0 + \psi_1 z.
\end{equation}
Each element of the Hilbert space ${\cal H}$
 will be then characterized by two complex numbers.
 We have recovered the
standard SU(2) two-component spinors.
The inner product is then,
\begin{equation}
\langle \Phi | \Psi\rangle = \frac{1}{2} (\bar{\phi}_0 \psi_0 +
\bar{\phi}_1 \psi_1) .
\end{equation}
For details see \cite{wood}.

\subsubsection{Geometry of a Quantum Spin-1/2 System}

The spin degrees of freedom of a spin 1/2 particle provide
a very clear example
of the geometric structures described in  Sec. ~\ref{sec:sec2a}.
In this case the Hilbert space ${\cal H}$ is formed by vectors on
$C^2$:$\left( \begin{array}{c} \alpha \\ \beta \end{array} \right)$
where $\alpha$ and $\beta$ are complex numbers.
As we saw above,
it is convenient then to consider ${\cal H}$ as a real vector space.
 Instead of a
column vector in $C^2$ we will have  column vectors on $R^4$:
\begin{equation}
\Psi = \left(\begin{array}{c} a \\ b\\ c \\ e \\ \end{array} \right),
\end{equation}
where $a,b,c,e$ are real numbers.

The Hermitian inner product $\langle \Psi |\Phi \rangle$ between
$\Psi=\left( \begin{array}{c} \alpha
 \\ \beta \end{array} \right)$ and
$\Phi=\left( \begin{array}{c} \gamma \\
\delta \end{array} \right)$
given by
\begin{equation}
\langle \Psi |\Phi \rangle =\bar{\alpha}\,\gamma
 + \bar{\beta}\,\delta
\end{equation}
induces a metric $G$ and a symplectic two form $\Omega$ on $R^4$:
\begin{eqnarray}
G_{ij} &=& 2\left[ \nabla_i(a)\nabla_j(a) + \nabla_i(b)\nabla_j(b)
 +\nabla_i(c)\nabla_j(c) + \nabla_i(e)\nabla_j(e)\right],
 \nonumber \\
\Omega_{ij} &=& 4\left( \nabla_{[i}a\nabla_{j]}b +
\nabla_{[i}c\nabla_{j]}e\right).
\end{eqnarray}
Normalized states satisfy then,
\begin{equation}
\langle\Phi | \Phi\rangle = a^2 + b^2 + c^2 + e^2 =1.
\end{equation}
Thus, the space $S$ corresponds to the 3-sphere $S^3$.

We know that the quantum space of states ${\cal P}$
 will be the projection of
$S^3$ under the action of the $U(1)$ action.
 That is, $S$ has the structure of
a principal fiber bundle with fiber $S^1$
 and base space ${\cal P} = S^2$:
\begin{eqnarray}
S^1 \longrightarrow &S&^3 \nonumber \\
                  \pi  &\downarrow&\nonumber \\
&S&^2
\end{eqnarray}
This corresponds precisely to one of the Hopf
 bundles over the two sphere $S^2$.

In order to show the projection $\pi$
 explicitly and recover common
 coordinates on
the sphere $S^2$ we introduce the  coordinates
 $(\alpha, \beta, \delta)$ on $S^3$ as
follows,
\begin{eqnarray}
a &=& \cos(\textstyle\frac{\beta}{2})
 \cos(\delta + \alpha),\nonumber \\
b &=& \cos(\textstyle\frac{\beta}{2})
 \sin(\delta + \alpha),\nonumber \\
c &=& \sin(\textstyle\frac{\beta}{2})
 \cos(\delta - \alpha),\nonumber \\
e &=& \sin(\textstyle\frac{\beta}{2})
 \sin(\delta - \alpha).\nonumber \\
\end{eqnarray}
It is straightforward to compute the
induced simplectic structure on $S$:
\begin{equation}
\bar{\Omega}_{ij} =4 \sin \beta
 \nabla_{[i}\alpha \nabla_{j]}\beta.
\end{equation}
 It is clear that the degenerate direction of $\bar{\Omega}$ is
 $\left(\frac{\partial}
{\partial \delta}\right)^j$,
which is precisely the direction of the
 `phase change' generated by $J$.

The induced metric on $S$ is
\begin{equation}
\bar{G}_{ij} = \nabla_i (\alpha) \nabla_j (\alpha) +\frac{1}{4}
 \nabla_i (\beta) \nabla_j (\beta)
 +\nabla_i (\delta) \nabla_j (\delta)-
2\cos\beta \nabla_{(i} (\alpha) \nabla_{j)} (\delta).
\end{equation}
It is clear that  $\bar{\Omega}$  correspond to the pullback of
 ${\bf \Omega}$ under $\pi$ ($\bar{\Omega}=
 \pi^{\ast}{\bf \Omega}$). We can
find the metric defined in the orbits of
 the degenerate direction, and
define $(g, {\bf \Omega})$ on ${\cal P} = S^2$ with
ordinary spherical coordinates $(\theta=\beta,
 \phi=2\alpha)$ to be
\begin{eqnarray}
{\bf \Omega}_{ab} & = &2 \sin \theta
 \nabla_{[a}\phi \nabla_{b]}\theta,
 \label{3.21}\\
g_{ab} & = &\frac{1}{2}\left[ \sin^2 (\theta)\,
 \nabla_a (\phi) \nabla_b (\phi) +
 \nabla_a (\theta) \nabla_b (\theta)\right].\label{3.22}
\end{eqnarray}

Quantum observables correspond on ${\cal H}$ to Hermitian
 $2\times 2$ matrices.
A basis for those matrices is given by the Pauli matrices.
 They are associated with the
generators of rotations in 3 dimensions and are the
`angular momentum' operators
 $\hat{S}_x,\,\hat{S}_y$ and $\hat{S}_z$, satisfying ordinary
 commutation relations:
$[\hat{S}_i, \hat{S}_j]=\hbar \varepsilon_{ijk}\hat{S}_k$.
 We know that there are three functions on ${\cal P}$
 which correspond to the
`observables' in the `quantum phase space';
\begin{eqnarray}
x:=\langle \hat{S}_x \rangle & = & \hbar (a\,c + b\,e ) =
\textstyle\frac{\hbar}{2} \sin\theta\,
\cos \phi ,\nonumber \\
y := \langle \hat{S}_y \rangle & = & \hbar (a\, e - c\, b)  =
 \textstyle\frac{\hbar}{2} \sin\theta\,
\sin \phi ,\nonumber \\
z := \langle \hat{S}_z \rangle & = &
 \textstyle\frac{\hbar}{2}\left[ (a^2 + b^2) -
 (c^2 + e^2)\right]
= \textstyle\frac{\hbar}{2} \cos \theta.
\end{eqnarray}
It is a curious fact that they are also
 the coordinates of a
sphere of radius $\hbar/2$.

Let us now consider dynamical evolution.
 Without loss of generality we can
take the Hamiltonian to be $H=A \hat{S}_z$. The corresponding
 observable on ${\cal P}$
is $h=\langle \hat{H} \rangle=A\,
\textstyle\frac{\hbar}{2} \cos\theta$.
Given $h$ and ${\bf \Omega}$ we can compute
 the equations of motion for the
coordinates $(\theta, \phi)$:
\begin{eqnarray}
\dot{\theta}&=&{\bf \Omega}^{ab}\,
\partial_a\theta \,\partial_b h = 0
 ,\nonumber \\
\dot{\phi} &=& {\bf \Omega}^{ab}\,\partial_a\phi
\,\partial_b h=-A
\textstyle\frac{\hbar}{2}.
\end{eqnarray}
That is, the quantum evolution is given by a
`point' traveling on $S^2$
 at constant
`latitude' $\theta$ and with constant angular velocity
$\dot{\phi}=-A\textstyle\frac{\hbar}{2}$.

Note that the quantum description in terms of
 `K\"ahler geometry' for
the spin-1/2 particle
coincides exactly with the classical
 description given in Sec.~\ref{sec:sec2}.
for the chosen
Hamiltonian. The spheres in both cases have, however,
 very different origin. In
one case it is the smallest quantizable
 {\it reduced phase space}. In the quantum
case is the {\it projective\/} `quantum phase space'
 coming from the Hilbert
space of states.

\section{Nonstandard Quantum Hamiltonian Systems}
\label{sec:sec4}

Notice that our previous discussion means
 that it is possible to describe
the quantization of a system in two stages.
In order to see this, it is
simpler to think of these stages in reverse,
 that is, as one method of
constructing a classical theory from a known
 quantum theory.  In this
`classicalization' one would begin
 with a Hilbert space ${\cal H}$ and a
set of observables given as linear
operators on ${\cal H}$.  We could now
project to the space of rays ${\cal P}$, which,
 since it is a phase space
itself and observables are now represented
 by real valued functions, the
system is represented by a `classical theory'
 with at least a large part
(if not all) of the content of the quantum
 theory defined on the Hilbert
space.  The main addition to this
 `classical' theory is the probability
structure given by (\ref{3.7}) based on the
 Riemannian metric $g_{ab}$. If one
were able to ignore the probability structure
 of this symplectic
manifold, one could think of quantum mechanics
on ${\cal P}$ as nothing
more than another classical theory.  Our program
of `classicalization'
would then be simply a map from ${\cal P}$ to
another symplectic manifold
$\Gamma$, the phase space of the usual classical
 theory.  We can
represent the process by the following diagram,
\begin{equation}
\matrix{{\cal H}&&\cr \downarrow&&\cr {\cal P}&
 \rightarrow & \Gamma\cr}
\label{4.1}
\end{equation}

The usual process of `quantization' is to
 leap from $\Gamma$ directly to
${\cal H}$, but one might just try to reverse
 the direction of the arrows in
(\ref{4.1}), first constructing the `K\"ahler
 quantum theory' on ${\cal P}$,
then `raising' the observables on ${\cal P}$
 to Hermitian operators on
${\cal H}$.  Notice that it could be
 possible to stop this procedure at
${\cal P}$ if one could be certain that {\it all\/} the
properties of quantum mechanics (such as
the superposition of states)
could be realized in terms of
 observables on ${\cal P}$ and the
probability structure generated by $g_{ab}$.

The program we are addressing in this
paper involves, however, the
ordinary quantization process from $\Gamma$
 to ${\cal H}$ and then considering
the `projected' geometrical formulation on ${\cal P}$.
 The classical theory we are
starting with, having a modified
 symplectic geometry defined on it, will
yield a different geometry on ${\cal P}$.
 That is, the symplectic
structure ${\bf \Omega}$ on ${\cal P}$
 will have some information of the
corresponding one on $\Gamma$.
 The question we are led to ask is: Is the
`non-standard' geometry induced on the
constructed quantum theory compatible
 with experiment?

 From now on we will restrict our attention
 to the spin-1/2 system, and show
explicitly that there are obstructions at
 each level to this procedure.
Given that the various Hamiltonian
 descriptions for the classical system
differ by only a conformal transformation,
 the set of issues we will be
addressing are the ones we called `dynamical'
 in the discussion at the beginning
 of Sec.~\ref{sec:sec3}.
While we will see that it is quite simple
to mirror the change of
symplectic structure given by (2.15) and
recover the dynamics of the
quantum system on ${\cal P}$ (in the sense
of recovering the integral
curves of the original system), but we will
 find that it is more
difficult to maintain the probability
 structure in terms of $g_{ab}$
that does not exist in the purely classical system.
 We will also see
that realizing the dynamics of the
 nonstandard Hamiltonian system in
terms of a linear Hamiltonian operator is
impossible in most cases.

We would like to change the symplectic two-form
on ${\cal P}$ for the
spin-1/2 system and find a new
Hamiltonian function $\tilde h$ which
gives the same set of integral curves
 that are given in Sec.~\ref{sec:sec3}.
We must also require that the {\it physical\/}
 predictions be the same in terms of
measurement. Recall that the probability of
 measuring the eigenvalue $o_i$ of an
operator $\hat{O}$ when the system is in
 state $\Psi$ is given by the geodesic
distance from $\Psi$ to the point $\Phi_i$
 ($\hat{O} \Phi_i=o_i \Phi_i$):
$P(\Psi, o_i) = \cos^2\left[\frac{d(\Psi,\Phi_i)}
{\sqrt{2}}\right]$.
 This implies that in
order to recover the same physical predictions,
 not only the dynamical
trajectory must
be the same but also the geodesic distance to
 the eigenstates.

Let us consider a double Stern-Gerlach
 experiment in which we first measure
$\hat{S}_z$ and
then look only at the particles that had
spin `up'. In our picture, this corresponds
to considering a quantum state located at the
`north pole' ($\theta=0$). We put now
a second measuring device. The spatial orientation of
the apparatus corresponds
precisely to the orientation of the eigenstates
 (which lie on `antipodal points')
on the sphere. The probability of measuring spin `up'
 and `down' will depend only
on the angle along maximal circles, from the north
 pole to the `podes'.
 Since the system is rotationally symmetric, we
can rotate both detectors while keeping their
 relative orientation fixed and the
probabilities will not change. That operation corresponds to
`fixing the `up' direction
of the detectors' in $(x,y,z)$ space
and rotating  the sphere. Since the
distance along the sphere must be the same, we conclude
 that the metric on $S^2$ should be
rotational symmetric, which is a property of the
 metric inherited from
the Hermitian inner product. Let us denote  by
$\stackrel{o}{g}_{ab}$, the
metric defined by Eq. (\ref{3.22})
 $\left( \stackrel{o}{g}_{ab}=\frac{1}{2}\left[
\sin^2 (\theta)\, \nabla_a (\phi) \nabla_b (\phi) +
 \nabla_a (\theta) \nabla_b (\theta)\right]\right)$.

We can conclude then that the metric $g$ should be
 equal to $\stackrel{o}{g}$,
 together with the
integral curves. The question that we are led to
 ask is: can we find a new
 $\tilde{\Omega}$ and $\tilde{h}$ such
that the Hamiltonian vector field of $\tilde{h}$
 and $g_{ab}$ are the same? Since any two-form on $S^2$
 is given by a conformal
transformation from the `canonical'
two-form ${\bf \Omega}$ defined by eq. (\ref{3.21}),
what we are looking for is precisely the
 conformal factor $\mu$
 in Sec.~\ref{sec:sec2}. such that,
\begin{equation}
\tilde{\Omega}^{ab}=\mu {\bf \Omega}^{ab}.
\end{equation}

It is easy to see that we can find a $\tilde{h}$
such that the dynamical evolution
is the same. The condition, in the $(\theta, \phi)$
coordinates, is
\begin{equation}
\left(\begin{array}{c} 0 \\ -A \textstyle\frac{\hbar}{2}
 \end{array}\right)=
\left(\begin{array}{cc} 0 & \tilde{\Omega}^{\theta \phi}
\\
-\tilde{\Omega}^{\theta \phi} & 0 \end{array}\right)
\left(\begin{array}{c}
\partial_\theta \tilde{h}\\ \partial_\phi \tilde{h}
 \end{array}\right).
\end{equation}
This set implies that $\partial_{\phi} \tilde h = 0$,
 or, $\tilde{h}=
f(\theta)$, so the system reduces to one equation:
\begin{equation}
 A \textstyle\frac{\hbar}{2} = \tilde{\Omega}^{\theta
 \phi} f^\prime,
\end{equation}
where $f^\prime=\frac{d f}{d\theta}$.

Therefore, $\tilde{\Omega}^{\theta \phi} =A
 \textstyle\frac{\hbar}{2}
 \frac{1}{f^\prime}$.
To solve the system, we could
fix $f$ and then define $\tilde\Omega$ from the previous equation.
 This would give us the
conformal factor to be $\mu =
 \frac{-\hbar A}{2}\frac{ \sin \theta}{f^\prime}$.

 However, recall
that ${\cal P}$ must have a K\"ahler structure, so $g$ and
 $\Omega$ must be compatible
in the sense that $g_{ab}=J^c_a \Omega_{cb}$. Can we change
 $\Omega$ arbitrarily
and still have a compatible system for fixed $g$?
The answer to this question is no.
A little lemma follows:
\newtheorem{lemma}{Lemma}
\begin{lemma}
Let $\stackrel{o}{g}_{ab}$ be the metric on $S^2$ given by
 Eq. (\ref{3.21}), then
$({\cal P}, g, {\bf \Omega})$ is a K\"ahler
 Manifold iff $f^\prime=K \sin\theta$.
That is iff $\mu=C$, where $K$ and $C$ are real constants.
\end{lemma}
We have to conclude, that it is impossible to have
 a nonstandard quantum Hamiltonian
dynamics compatible with observation: there is no
 freedom in changing $\Omega$ and $h$.

The second obstruction (the two obstructions are
 probably strongly
related) to changing
the symplectic structure in quantum mechanics
is that we would normally
like
to have the `K\"ahler quantum mechanics' on
 ${\cal P}$ come from
a system of operators in a Hilbert space whose
expectation values on ${\cal P}$ would
 generate the observables.
If we attempt to do this for $\tilde h$, and even
 if we were to ignore the
lemma above, we are
still restricted by the fact  that $\tilde h$ must be a
function of $\theta$ only.  Even if we try to let
 $\tilde h$ be
any function of $\theta$, in this simple case if
 $\tilde h$ is to be the
image of
a linear Hermitian operator on the space of vectors
 in ${\cal H}$,
the operator $\hat {\tilde H}$ must be of the form
\begin{equation}
\hat {\tilde H} = \zeta I +
\textstyle\frac{\eta}{2}\hat S_x +
\textstyle\frac{\kappa}{2} \hat S_y +
 \textstyle\frac{\lambda}{2} \hat S_z,
\end{equation}
with $\zeta$, $\eta$, $\kappa$, $\lambda$ real.
 This means that
\begin{eqnarray}
\tilde h & = & \zeta + \textstyle\frac{\eta}{2}
\langle\hat S_x \rangle +
\textstyle\frac{\kappa}{2} \langle\hat S_y\rangle
 + \textstyle\frac{\lambda}{2}\langle\hat S_z\rangle
\nonumber\\
& = & \zeta + \eta \textstyle\frac{\hbar}{4}
\sin \theta \cos \phi  + \kappa
\textstyle\frac{\hbar}{4} \sin \theta \sin \phi +
\lambda \textstyle\frac{\hbar}{4}
 \cos \theta
\end{eqnarray}
must be a function of $\theta$.  The only way to
satisfy this for all $\phi$
is to take $\eta = \kappa = 0$.  This
means that the only possible $\tilde h$ that come
 from linear Hermitian
operators are
\begin{equation}
 \tilde h = Kh + D,
\end{equation}
where $K$ and $D$ are real constants.
 In this case the new $\mu$ is $\mu
= (1/K)\mu_0$.  All other choices of
 $\mu$ must lead to $\hat {\tilde H}$ a
nonlinear operator.

\section{Conclusions and Suggestions For Further Research}
\label{sec:sec5}

We have attempted to transfer to quantum
 theory an idea originally due to
Hojman, that perhaps the usual
symplectic structure of classical mechanics
is too restrictive, and it
might be possible to generalize it.
 In classical mechanics this is
certainly the case, and it may lead
 to new approaches to solving old
problems, and can be used to construct
 Hamiltonian theories for systems
that have no variational principle, and
 thus no Hamiltonian in the usual
sense.  We have considered this idea from
 the viewpoint of changing the
symplectic structure and Hamiltonian of a
 system that does have a
Hamiltonian.  Classically this can be done
with no loss of generality,
since we can easily generate the same solution
 curves for the system for
a large class of symplectic structures.

What we have just shown is that,
even in the  Ashtekar-Schilling formulation \cite{GQM2},
 where the evolution of
the system takes place on a symplectic
 manifold similar to that of
classical mechanics, the extra rigidity a
 probability structure imposes
on the system makes it impossible to
 use symplectic structures of the
type we have been able to use in classical mechanics.
 In fact, our
spin-1/2 example shows that the restrictions
 on the symplectic structure
are quite strong.  A probably related obstruction is
 that only certain
Hamiltonians on ${\cal P}$ can be represented
 by linear Hermitian operators on the
Hilbert space ${\cal H}$ that generates the symplectic
 manifold ${\cal P}$.

It seems, then, that the results of the
article are essentially
negative.  However, it may be possible to change
 some of the structures
on the quantum symplectic manifold in order to try
 to maintain the idea
of a more general symplectic structure while
 still keeping the
probability structure necessary for quantum
 mechanics.

There are two obstructions to the program of
generalizing symplectic
structures.  Perhaps the most important is the
 fact that changing
$\Omega_{ab}$ on ${\cal P}$ leads to a disastrous
 change in the metric
$g_{ab}$ on ${\cal P}$ that defines the probability.
  If it were possible
to change $\Omega_{ab}$ without changing $g_{ab}$,
 we would have a simple
solution to the problem.  The difficulty here is
 Eq. (\ref{1.10}),
\begin{equation}
g_{ab} = \Omega_{ac}J^c_b,\nonumber
\end{equation}
which relates $\Omega_{ab}$ to $g_{ab}$
 through the complex structure
tensor $J^a_b$.  Note that the complex
 structure is required to obey
$J^b_a J_b^c = -\delta^c_a$.  If we make
a similarity transformation
(such as a coordinate transformation) on $J$,
 $J^a_b = S^a_c J^c_d
(S^{-1})^d_b$, $J^b_a J^c_b = -\delta^c_a$ is preserved.
  If one makes
such a transformation, both $\Omega_{ab}$ and $g_{ab}$ change
as `covariant tensors', which is perfectly acceptable.
  Notice that if we
were to make a more complicated
transformation, such as a conformal
transformation, on $\Omega_{ab}$,
 $\Omega_{ab} \rightarrow \varphi
\Omega_{ab}$, and at the same time
insist that $g_{ab}$ remain unchanged
in order to preserve the probability structure,
 we would have to allow
$J^a_b \rightarrow (1/\varphi)J^a_b$,
 and $J^b_a J^c_b =- (1/\varphi)^2
\delta^c_a$, which is negative definite
 and nonsingular as long as
$\varphi$ is finite and nonzero, but
 does not obey the defining
equation of a complex structure tensor.
 We have been unable to find in
the literature any study of this type of
`pseudocomplex structures'
which would allow more drastic changes
 in $J^a_b$, and it might be
worthwhile to consider such objects
 to see if a consistent quantum
mechanics on ${\cal P}$ could be constructed using them.
  We have taken a
conformal transformation as an example
 because in our spin-1/2 system,
with its low-dimensional phase space,
 the Hojman transformation (\ref{hoj})
 reduces to a simple conformal transformation.

In higher dimensional phase spaces the
 Hojman transformation $\Omega_{ab}
\rightarrow K^c_a \Omega_{cb}$ would imply
that to maintain the metric
$g_{ab}$ invariant one would have to
take $J^a_b \rightarrow J^{\prime
c}_b = J^a_c (K^{-1})^c_b$, and, in principle,
 since the Hojman
transformation contains the conformal
 factor $\mu$, we might expect that
$J^{\prime a}_b J^{\prime b}_c$ would
 not be equal to $-\delta^a_c$, just
as for a conformal transformation.
 In that case, it would be
necessary to postulate `pseudocomplex
structures' similar to those just
mentioned in order to preserve $g_{ab}$
 on changing $\Omega_{ab}$.  Note,
however, that while the Hojman
 transformation for a two-dimensional phase
space reduces to a pure conformal transformation,
 the more general
transformation allowed in higher dimensional
 phase spaces may still allow
us to write $J^{\prime a}_b J^{\prime b}_c =
-\delta^a_c$, in which case
$J^{\prime a}_b$ is nothing more than a
`deformed complex structure',
and this concept has been studied for
 some time \cite{defcom}.
 It is necessary to
investigate whether the Hojman transformation
allows $J^{\prime a}_b
J^{\prime b}_c = -\delta^a_c$ or not.

Another possibility that would allow change
 in the symplectic structure
without deforming the complex structure
 would be to allow the appropriate
transformation on $g_{ab}$ that would
 preserve $J^a_b$ (in the spin-1/2
case a conformal transformation) and
 define probabilities in some
`conformally invariant' fashion.
 We will not attempt to consider this
idea further.

One remark is in order. The phase space of
the system we started with,
 namely
the sphere $S^2$, is somewhat special.
 Perhaps the most notorious
property is that it is a {\it compact\/}
manifold. As a consequence, the
Hilbert space in the quantum theory is
{\it finite\/} dimensional. Furthermore,
 it has
recently been shown that the {\it only\/} classical
observables that can be
quantized in a way that the prescription $\{ ,\}
\rightarrow i\hbar[,]$
is satisfied exactly, are the generators of
 rotations $s_i$ \cite{gotay}.
This is the equivalent, for $S^2$, of the
 Groenewold-Van Hove
 theorem\cite{gron}.
Our result for the spin-1/2 system is therefore another
indication of the
`rigidity' of the structures one can define on the sphere.
This has to be contrasted with higher
 dimensional (non-compact) phase spaces
for which the quantum theory is much richer
(infinite dimensional Hilbert
 space). In this case one has in fact an infinite
 number of possible
complex structures (this freedom is similar to the
 one that leads to different
inequivalent  representations of the CCR in QFT).
In this case, the
nonstandard quantum theory has to satisfy the `kinematical'
 requirements related
to the Heisenberg uncertainty principle,
 and possible super-selected sectors, in
order to be considered `valid'. A complete study of
the most general case is
 therefore,  still open.

Finally, note that if it were possible to be sure that
 all of the content of
quantum mechanics could be achieved in terms of the
evolution and
structure of points in ${\cal P}$, we would not need
 to worry about the
fact that the time evolution of states,
for example, is a reflection of
evolution in the Hilbert space ${\cal H}$
  that is generated by a
nonlinear Hamiltonian operator.
 If this is not so, then we would be
forced to consider the possibility of nonlinear
 evolution in quantum
mechanics, an idea that has been proposed by several
 authors \cite{nonl}, but one
should be justifiably reluctant to propose such
 a drastic modification
to, at the very least, a one-particle model.

\acknowledgments

We would like to thank S. Hojman and A. Gomberoff
 for stimulating discussions.
We also thank T. Schilling for discussing his work with us,
and J.A. Zapata for helpful comments.
 One of us (MR) is
grateful to the Center for Gravitational Physics
 and Geometry for the
hospitality during 1994-95.
Both authors were supported by the National University
of M\'exico (DGAPA-UNAM).
This work was supported in part by the NSF
grant PHY93-96246, by the Eberly
research fund of Penn State University, and by a
 CONACYT-CONICET exchange grant
(M\'exico-Chile).

\end{document}